\documentclass[reqno,12pt]{article}
\usepackage{amsmath,amsfonts,amssymb,amsthm,amstext,amscd,eucal}
\usepackage{amsmath,multirow,latexsym}
\usepackage{srcltx}
\usepackage{color}
\newcommand{\be}{\begin{equation}}
\newcommand{\ee}{\end{equation}}
\newcommand{\bn}{\begin{eqnarray}}
\newcommand{\en}{\end{eqnarray}}
\newcommand{\bnn}{\begin{eqnarray*}}
\newcommand{\enn}{\end{eqnarray*}}
\newcommand{\ba}{\begin{array}}
\newcommand{\ea}{\end{array}}

\newcommand{\R}{{\mathbb R}}

\newcommand{\biz}{\begin{itemize}}
\newcommand{\eiz}{\end{itemize}}
\newcommand{\ben}{\begin{enumerate}}
\newcommand{\een}{\end{enumerate}}

\newcommand{\vfx}{\varphi\,(x) }

\newcommand{\vff}{\varphi_f }

\newcommand{\vffkp}{\varphi_{f_k}\,  \cdots  \,\varphi_{f_1}}

\newcommand{\vffkpop}{\varphi_{f_{k + 1}}\,  \cdots  \,\varphi_{f_n}}

\newcommand{\wf} {W\,(x_1,  \ldots, x_n)}

\newcommand{\fsx}{f\,(x)}

\newcommand{\fsy}{f\,(y)}

\newcommand{\fsbxsp}{\overline{f_1\,(x_1)}\,\star  \cdots \star \overline{f_k\,(x_k)}
\star f_{k + 1}\,(x_{k + 1})\,\star\,  \cdots \star \,{f_n\,(x_n)}}

\newcommand{\dxo}{d\,x_1}

\newcommand{\dxn}{d\,x_n}

\def\nc{noncommutative }
\def\ncy{noncommutativity }
\def\th{$\theta_{\mu\nu}$}

\def\ss{$\theta_{0i}=0$}
\def\group{${\cal P}(1,1)\times E_2$}

\begin{document}

\begin{center}
{{\Large {\bf  Towards an Axiomatic Formulation\\\vskip0.5cm of
Noncommutative Quantum Field Theory
}}}%
\vspace*{10mm}

{\bf \large{ M. Chaichian$^\dagger$, M. N. Mnatsakanova$^{\S}$,
\begin{tabular}{|c|}\hline K. Nishijima$^\ddag$\\ \hline
\end{tabular}, A. Tureanu$^\dagger$ and  Yu. S.
Vernov$^\P$}}

\vspace*{7mm}

{\it $^\dagger$Department of Physics, University of Helsinki,
P.O. Box 64, 00014 Helsinki, Finland\\
$^{\S}$ Skobeltsyn Institute of Nuclear Physics, Moscow State University,\\ 119992, Vorobyevy Gory, Moscow, Russia\\
$^\ddag$Department of Physics, University of Tokyo 7-3-1 Hongo,
Bunkyo-ku,\\ Tokyo 113-0033, Japan\\
$^{\P}$Institute for Nuclear Research, Russian Academy of
Sciences,\\
60th October Anniversary prospect 7a, 117312, Moscow, Russia}

\end{center}

\abstract{We propose new Wightman functions as vacuum expectation
values of products of field operators in the \nc space-time. These
Wightman functions involve the $\star$-product among the fields,
compatible with the twisted Poincar\'e symmetry of the
noncommutative quantum field theory (NC QFT). In the case of only
space-space \ncy ($\theta_{0i}=0$), we prove the CPT theorem using
the \nc form of the Wightman functions. We also
show that the spin-statistics theorem, demonstrated for the simplest
case of a scalar field, holds in NC QFT within this formalism.}


\section{Introduction}
The axiomatic approach to quantum field theory (QFT) built up by
Wightman, Jost, Bogoliubov, Haag and others made QFT a consistent,
rigorous theory (for references, see Refs. \cite{WS}-\cite{Haag}). In the
framework of this approach, fundamental results, such as the CPT and
spin-statistics theorems, were proven in general, without any
reference to a specific theory and a Lagrangian or Hamiltonian. In
addition, the axiomatic formulation of QFT has given the possibility
to derive analytical properties of scattering amplitudes and, as a
result, dispersion relations. Consequently, various rigorous bounds
on the high-energy behaviour of scattering amplitudes were obtained
\cite{Froissart}.

At present, \nc quantum field theory (NC QFT) attracts a great deal
of attention. The study of such theories has received a considerable
impetus after it was shown that they appear naturally, in some
cases, as low-energy limit of open string theory in the presence of
a constant antisymmetric background field \cite{SW}. In this
context, the coordinate operators of a space-time satisfy the
Heisenberg-like commutation relations
\be\label{cr} [\hat{x}_\mu,\hat{x}_\nu]=i\theta_{\mu\nu}, \ee
where $\theta_{\mu\nu}$ is a constant antisymmetric matrix of
dimension (length)$^2$. For the study of NC QFT it is customary to
define for the field operators on the noncommutative space-time,
$\phi(\hat x)$, their Weyl symbols, $\phi(x)$, whose algebra is
isomorphic to the initial operator algebra. The connection between
the operators $\phi(\hat x)$ and their Weyl symbols $\phi(x)$ is
achieved through the Weyl-Moyal correspondence, which requires that
products of operators are replaced by Moyal $\star$-products of
their Weyl symbols: \be\label{corres} \phi(\hat x)\psi(\hat
x)\rightarrow \phi(x)\star\psi(x), \ee where the Moyal
$\star$-product is defined as
\be\label{star} (\phi\star\psi)(x)=\phi(x)\
e^{\frac{i}{2}\theta^{\mu\nu}\overleftarrow{\partial_\mu}\overrightarrow{\partial_\nu}}\
\psi(x)\ . \ee

In the following, by field operators we shall understand such Weyl
symbols. Since the matrix $\theta_{\mu\nu}$ is constant and does not
transform under Lorentz transformations, the Lorentz symmetry
$SO(1,3)$ is broken, while the translational invariance is
preserved.

We shall consider throughout this paper only the case of space-space
noncommutativity, i.e. $\theta_{0i}=0$, since theories with
$\theta_{0i}\neq 0$ cannot be obtained as low-energy limit from
string theory \cite{SW}. Besides, it has been shown that such field
theories violate perturbative unitarity \cite{GM} and causality
\cite{SSB,cnt}.

The study of NC QFT has been mostly done in the Lagrangian approach
(for reviews, see Refs. \cite{DN,Szabo}). However, it would be of
importance to develop also an axiomatic formulation of NC QFT, which
does not refer to a specific Lagrangian.

The first step in this direction was made in Ref. \cite{LAG}, where the
usual Wightman functions, defined as
 \be\label{wightman} W(x_1,
x_2,...,x_n)=\langle0|\phi(x_1)\phi(x_2)...\phi(x_n)|0\rangle \ee
were investigated, but based on the symmetry group $O(1,1)\times
SO(2)$, which is the residual space-time symmetry of NC space-time
with \ss. Using the usual Wightman functions as in (\ref{wightman}),
the validity of the CPT theorem was shown in Ref. \cite{LAG}.

Perhaps the most serious problem with NC QFT treated in the residual
symmetry approach, used to be the fact that the representations of
the fields did not match the actual symmetry of the theory. The
groups $O(1,1)$ and $SO(2)$ are both Abelian, having only
one-dimensional irreducible representation, and therefore not
supporting the concept of spin, which is essential for particle
physics. A rigurous axiomatic approach should include the proof of
the spin-statistics theorem, but the spin simply does not exist in a
$O(1,1)\times SO(2)$-invariant theory, as pointed out in Ref. \cite{LAG}.
The solution to the representations problem was the main physical
implication of the uncovering of the twisted Poincar\'e symmetry of
noncommutative quantum field theory with Heisenberg-like commutation
relations of the coordinates \cite{CKNT}. Twisted Poincar\'e
symmetry proved to be the new concept of relativistic invariance for
NC QFT \cite{CPrT} and it shaped the field in a more rigorous way.
Based on the twisted Poincar\'e symmetry and its various
consequences, we are now in a position to formulate a
well-argumented axiomatic approach to NC QFT.

The role of twisted Poincar\'e symmetry in this paper is two-fold.
Firstly, it is needed in order to justify the use of the
star-product between functions in two different space-time points,
since in this case the newly defined Wightman functions will have
the same symmetry as the one of space-time commutation relation. In
particular, for scalar fields the new Wightman functions will be
explicitly scalars under the twisted Poincar\'e transformations.
Secondly, since the Lorentz invariance is violated in such NC
theories down to the product of two Abelian groups, one has not the
concept of spin and thus cannot speak of the spin-statistics theorem
altogether, unless one invokes the existence of twisted Poincar\'e
thanks to its representation theory being identical with the one of
the usual Poincar\'e symmetry.

The existence of the class of test functions in the case of NC
quantum field theories (see Ref. \cite{test_funct}), is
crucial for utilizing the analytical properties of the smeared new
Wightman functions needed for the rigorous proof of the CPT and
spin-statistics theorems, as it is the case also for the usual
commutative quantum field theories. We would like to mention, however, that the field operators, even in the free case, with usual perturbative quantization procedure, do not satisfy the deformed locality condition nor the weak locality condition (see Section \ref{microcausality}).

In this paper, we shall formulate the axiomatic approach to NC QFT
mainly guided by the twisted Poincar\'e symmetry. The same symmetry
arguments impose also the adoption of a new form for the Wightman
functions. On the ground of this coherent formulation we shall prove
the CPT theorem for theories with space-space \ncy and also give the
proof of the spin-statistics theorem for the case of a spinless
field, for simplicity.

\section{Axiomatic approach to NC QFT}

\subsection{Twisted Poincar\'e symmetry}\label{twistPs}

Since the twisted Poincar\'e symmetry \cite{CKNT,CPrT} is our
guiding line in this formulation, we shall review a few main
concepts and formulas, for the consistency of the argumentation.

The twisted Poincar\'e algebra is the universal enveloping of the
Poincar\'e algebra $\cal U(\cal P)$, viewed as a Hopf algebra,
deformed with the Abelian twist element \cite{drinfeld} (see also
the monographs \cite{monographs})
\be\label{abelian twist}{\cal
F}=\exp\left({\frac{i}{2}\theta^{\mu\nu}P_\mu\otimes
P_\nu}\right),\ee
where $\theta_{\mu\nu}$ is a constant antisymmetric matrix (it does
not transform under the Lorentz transformations) and $P_\mu$ are the
translation generators. This induces on the algebra of
representations of the Poincar\'e algebra the deformed
multiplication,
\be\label{twist prod} m\circ(\phi\otimes\psi)=\phi\psi\rightarrow
m_\star\circ(\phi\otimes\psi)=m\circ{\cal
F}^{-1}(\phi\otimes\psi)\equiv \phi\star\psi\,,\ee
which is precisely the well-known Weyl-Moyal $\star$-product (taking
the Minkowski space realization of $P_\mu$, i.e.
$P_\mu=-i\partial_\mu$):
\be\label{star_prod}\star=\exp{\left(\frac{i}{2}\theta^{\mu\nu}\overleftarrow
\partial_\mu\overrightarrow\partial_\nu\right)}\ .\ee
In particular, taking in (\ref{twist prod}) $\phi(x)=x_\mu$ and
$\psi(x)=x_\nu$, one obtains:
\be\label{cr_symb} [x_\mu, x_\nu]_\star=i\theta_{\mu\nu}. \ee
This is the usual commutation relation of the Weyl symbols of the
noncommuting coordinate operators $\hat x$, \eqref{cr}, which is
obtained in the Weyl-Moyal correspondence.

The twist (\ref{abelian twist}) does not affect the actual
commutation relations of the generators of the Poincar\'e algebra
$\cal P$:
\bn [P_\mu,P_\nu]&=&0,\ \ \ \
[M_{\mu\nu},P_\alpha]=-i(\eta_{\mu\alpha}
P_\nu-\eta_{\nu\alpha}P_\mu),\cr
[M_
{\mu\nu},M_{\alpha\beta}]&=&-i(\eta_{\mu\alpha}M_{\nu\beta}-\eta_{\mu\beta}M_{\nu\alpha}-\eta_{\nu\alpha}M_{\mu\beta}+\eta_{\nu\beta}M_{\mu\alpha}).
 \en
Consequently also the Casimir operators remain the same and the
representations and classifications of \emph{particle states} are
identical to those of the ordinary Poincar\'e algebra. The question
of the fields, constructed by the method of induced representations,
is more subtle \cite{CKTZZ,CNST}, but crucial for the edification of
the axiomatic formulation on NC QFT, and we shall review it in the
next subsection.

The twist deforms the action of the generators in the tensor product
of representations -- the so-called {\it coproduct}. In the case of
the usual Poincar\'e algebra, the coproduct $\Delta_0\in\cal U(\cal
P)\times \cal U(\cal P)$ is symmetric (the usual Leibniz rule),
\be\label{primitive} \Delta_0(Y)=Y\otimes1+1\otimes Y,\ee
for all the generators $Y\in \cal P$. The twist $\cal F$ deforms the
coproduct $\Delta_0$ to $\Delta_t\in {\cal U}_t({\cal P})\times
{\cal U}_t(\cal P)$ as:
\be\label{twist_coproduct}\Delta_0(Y)\longmapsto\Delta_t(Y)={\cal
F}\Delta_0(Y){\cal F}^{-1}\,.\ee
This similarity transformation is compatible with all the properties
of $\cal U(\cal P)$ as a Hopf algebra, since $\cal F$ satisfies the
twist equation \cite{drinfeld}:
\be\label{twist_eq} {\cal F}_{12}(\Delta_0\otimes id){\cal F}={\cal
F}_{23}(id\otimes\Delta_0){\cal F}\,, \ee
where ${\cal F}_{12}={\cal F}\otimes 1$ and ${\cal F}_{23}=1\otimes
{\cal F}$. The eq. \eqref{twist_eq} ensures the associativity of the
$\star$-product \eqref{twist prod}. This is an important point, to
which we shall return when discussing the new form of the Wightman
functions.

The twisted coproducts of the generators of Poincar\'e algebra turn
out to be:
\bn \Delta_t(P_{\mu})&=& \Delta_0(P_{\mu})=P_{\mu}\otimes
1+1\otimes P_{\mu},\label{twist p}\\
\Delta_t(M_{\mu\nu})&=&M_{\mu\nu}\otimes 1+1\otimes
M_{\mu\nu}\label{twist
m}\\
&- &\frac{1}{2}\theta^{\alpha\beta}
\left[(\eta_{\alpha\mu}P_\nu-\eta_{\alpha\nu}P_\mu)\otimes
P_\beta+P_\alpha\otimes
(\eta_{\beta\mu}P_\nu-\eta_{\beta\nu}P_\mu)\right].\nonumber\en
Thus the twisted coproduct of the momentum generators is identical
to the primitive coproduct, eq. (\ref{twist p}), meaning that
translational invariance is preserved, while the twisted coproduct
of the Lorentz algebra generators, eq. (\ref{twist m}), is
nontrivial, implying the violation of Lorentz symmetry.

It is essential for our purpose to note that, by fixing conveniently
the frame of reference, the matrix \th\ takes a block diagonal form:
\begin{eqnarray}\label{theta}
\theta_{\mu\nu}=\left(
\begin{array}{cccc}
0 &\theta' & 0  & 0 \\
-\theta' & 0 & 0  & 0 \\
0 & 0  &0 & \theta \\
0 & 0  & -\theta & 0
\end{array}
\right).
\end{eqnarray}
This form emphasizes the stability group of the matrix \th, i.e.
$SO(1,1)\times SO(2)$, which becomes $O(1,1)\times SO(2)$ as soon as
$\theta'=0$ (space-space noncommutativity). One immediately notices
that the coproducts of the generators of $SO(1,1)$ and $SO(2)$, i.e.
$M_{01}$ and $M_{23}$, remain primitive in this frame of reference:
$\Delta_t(M_{01})=\Delta_0(M_{01})$ and
$\Delta_t(M_{23})=\Delta_0(M_{23})$, which is the mark of the
preservation of ordinary invariance under the corresponding Lorentz
transformations.

Although the formulation of the symmetry as twisted Poincar\'e
algebra is very useful for noting the solution to the representation
problem, it is important for physical application to understand how
the corresponding finite transformations act (see, e.g., Ref.
\cite{CKTZZ}). In the case of the twisted Poincar\'e algebra ${\cal
U}_t(\cal P)$, its dual is the algebra of function $F_\theta(G)$ on
the Poincar\'e group $G$ with deformed multiplication. The algebra
$F(G)$, dual to ordinary ${\cal U}(\cal P)$, is generated by the
elements $\mathbf{\Lambda}^{\mu}_{\nu}$ and $\mathbf{a}^\mu$, which
are complex-valued functions, such that when applied to suitable
elements of the Poincar\'e group, they would return the familiar
real-valued entries of the matrix of finite Lorentz transformations,
$\Lambda^\mu_\nu$, or the real-valued parameters of finite
translations, $a^\mu$. In the case of $F_\theta(G)$, the functions
$\mathbf a^\mu$ are no more complex-valued and the following
commutation relations are obtained by requiring the duality between
${\cal U}_t(\cal P)$ and $F_\theta(G)$:
\bn\label{finite translations}[\mathbf
{a}^\mu,\mathbf{a}^\nu]&=&i\theta^{\mu\nu}-i\mathbf{\Lambda}^\mu_\alpha\mathbf{\Lambda}^\nu_\beta\theta^{\alpha\beta}\,,\cr
\ \  [\mathbf{\Lambda}^\mu_\nu,
\mathbf{a}^\alpha]&=&[\mathbf{\Lambda}^\mu_\alpha,\mathbf{\Lambda}^\nu
_\beta]=0,\ \ \ \mathbf{\Lambda}^\mu_\alpha, \mathbf{a}^\mu \in
F_\theta (G)\,.\en
Again, one notices that, when applied to elements of the Lorentz
group which do not belong to the above mentioned $SO(1,1)$ or
$SO(2)$ stability subgroups, the eqs. \eqref{finite translations}
lead to noncommutative translations, whose physical interpretation
is elusive \cite{CNST}. Ordinarily, in $F(G)$, the functions
$\mathbf a^\mu$ applied to elements of Lorentz group vanish, but in
the case of $F_\theta(G)$ this peculiar result obstructs the
physical interpretation of finite twisted Poincar\'e transformations
of even the coordinates, and the problem becomes worse for the
transformation of fields.

It turns out that these problems are solved by the same stratagem
which will permit us to define noncommutative fields using the
method of induced representations.

\subsection{Noncommutative field operators}\label{NC_field_op}

One would be tempted to say that the construction of a NC quantum
field theory through the Weyl-Moyal correspondence is equivalent to
the procedure of redefining the multiplication of functions, so that
it is consistent with the twisted coproduct of the Poincar\'e
generators (\ref{twist_coproduct}) \cite{CKNT}.

However, the definition of noncommutative fields and the action of
the twisted Poincar\'e transformations on them is not a trivial one.
Ordinary relativistic fields are defined by the method of induced
representations. In the commutative setting, Minkowski space is
realized as the quotient of the Poincar\'e group by the Lorentz
group, and a classical field is a section of a vector bundle induced
by some representation of the Lorentz group. This construction does
not generalize to the noncommutative case, because the universal
enveloping algebra of the Lorentz Lie algebra is not a Hopf
subalgebra of the twisted Poincar\'e algebra. As a result, Minkowski
space $\R^{1,3}$, which in the commutative setting is realized as
the quotient of the Poincar\'e group $G$ by the Lorentz group $L$,
G/L , has no noncommutative analogue.

This can be intuitively seen if we recall that an ordinary
commutative field is typically written as
$$
 \Phi= f\otimes
v\,, \ \ \ f\in C^\infty(\R^{1,3})\,,\ \ \ v \in V\,,$$
where $C^\infty(\R^{1,3})$ is the set of smooth functions on
Minkowski space and $V$ is a Lorentz-module. The tensor product
requires that the action of the Lorentz generators on $\Phi$ be
taken with twisted coproduct. However, this implies an action of the
momentum generators $P_\mu$ on the representations of the Lorentz
group $v$, and such an action is simply not defined.

One proposal for bypassing this predicament is to consider $V$ a
Poincar\'e-module, with trivial action of the momentum generators
\cite{CKTZZ}. Another proposal -- which we shall adopt in this paper
-- is to maintain $V$ as a Lorentz module, but to forbid the
transformations which cannot go through \cite{CNST}. In this way, we
induce only the Lorentz transformations corresponding to the
stability group of \th, but the fields will carry representations of
the full Lorentz group; consequently, the particle spectrum of the
noncommutative quantum field theory with twisted Poincar\'e symmetry
will have the richness of the relativistic quantum field theory. At
the same time, the problem of noncommutative finite translations is
also solved.

We emphasize that the fact that only certain Lorentz transformations
are allowed on the noncommutative fields is a strong indication of
the Lorentz symmetry violation. The differences between ordinary and
noncommutative quantum fields are drastic and there is no way to
justify, based on the twisted Poincar\'e symmetry, the claim that
the noncommutative fields transform under all Lorentz
transformations as ordinary relativistic fields \cite{FW}.

\subsection{Noncommutative Wightman functions}\label{NCwight}

In the ordinary axiomatic field theory, Wightman functions are
defined as vacuum expectation values of products of relativistic
field operators, \be\label{wight} W(x_1,
x_2,...,x_n)=\langle0|\phi(x_1)\phi(x_2)...\phi(x_n)|0\rangle\,. \ee

The twist changes the multiplication in the algebra of
representation of the Poincar\'e algebra,  and we expect that this
should apply as well to Wightman functions. Indeed, the product of
field operators with independent arguments, $\phi(x_1)$ and $\phi
(x_2)$, in as far as the space-time dependence is concerned, is an
element of the tensor product of two copies of
$C_\theta^\infty(\R^{1,3})$, and the rule of multiplication in this
case is:
\bn\label{tensor_copies} (f_1\otimes 1)(1\otimes f_2)&=&f_1\otimes
f_2,\cr
(1\otimes f_2)(f_1\otimes 1)&=&({\cal R}_2f_1)\otimes({\cal
R}_1f_2)\,,\ \ \ \ f_1,f_2\in C_\theta^\infty(\R^{1,3}),
 \en
where ${\cal R} \in {\cal U}(\cal P)\otimes {\cal U}(\cal P)$ is the
universal $\cal R$-matrix, which relates by a similarity
transformation the twisted coproduct $\Delta_t$ and its opposite
$\Delta_t^{op}=\sigma\circ\Delta_t$,
\be\label{r-matrix}{\cal R}\Delta_t=\Delta_t^{op}{\cal R}\,.\ee
In the case of twisted Poincar\'e algebra with the twist
\eqref{abelian twist}, the expression of the ${\cal R}$-matrix is
\be {\cal R}={\cal F}_{21} {\cal F}^{-1}={\cal F}^{-2}\,.\ee
The property \eqref{tensor_copies} is encoded in the product of
functions with independent arguments as an extension of the
$\star$-product:
\be\label{starxy} f(x)\star g(y)=f
(x)e^{\frac{i}{2}\theta^{\mu\nu}\frac{\overleftarrow\partial}{\partial
x^\mu}\frac{\overrightarrow\partial}{\partial y^\nu}}g(y)\,. \ee
Such a generalization of the $\star$-product for noncoinciding
space-time points has been previously proposed and used in a
different context \cite{Szabo}.

This generalization can be also motivated by the fact that in the
commutation relation of coordinate operators,
$$
[\hat x_\mu,\hat x_\nu]=i\theta_{\mu\nu}\,,
$$
the labels by which we designate the coordinate operators are not
relevant, since the \ncy is only between different components of
space directions (see Ref. \cite{qshift}).

In this formulation, we propose for the noncommutative Wightman
functions the expression:
\be\label{wightstar} W_\star(x_1,
x_2,...,x_n)=\langle0|\phi_\theta(x_1)\star\phi_\theta(x_2)\star...\star\phi_\theta(x_n)|0\rangle\,.
\ee
In this expression and in the following, we shall denote the
noncommutative field operators by the subscript $_\theta$, in order
to emphasize once more their different properties from the ordinary
relativistic field operators. The $\star$-product of operators in
\eqref{wightstar},
\be
\phi_\theta(x_1)\star\phi_\theta(x_2)\star...\star\phi_\theta(x_n)=e^{\frac{i}{2}\theta^{\mu\nu}\sum\limits_{a<b}\frac{\partial}{\partial
x_a^\mu} \frac{\partial}{\partial
x_b^\nu}}\phi_\theta(x_1)\phi_\theta(x_2)...\phi_\theta(x_n)\ , \ee
is obviously associative and for the coinciding points
$x_1=x_2=...=x_n$ becomes identical to the multiple Moyal
$\star$-product.

\subsection{Translational invariance of the Wightman
functions}\label{translation}

The NC Wightman functions \eqref{wightstar} are translationally
invariant, as in ordinary relativistic QFT. This is obvious from the
fact that the coproduct of the translation generators is not
deformed, consequently the translations will act in an ordinary
manner on the NC Wightman functions. However, there are again
differences compared to the relativistic case. In relativistic QFT,
the differences of coordinates on which the ordinary Wightman
functions depend can be written as a four-vector,
$\xi^\mu_i=x^\mu_i-x^\mu_{i+1}$, since the functional form in all
four coordinates is the same, due to Lorentz symmetry. In the
noncommutative case, the Heisenberg fields $\phi_\theta(x)$ depend
on the $\theta$-matrix, and the same is valid for the
$\star$-product of fields entering the NC Wightman functions. The
Wightman functions are covariant only under the stability group
$O(1,1)\times SO(2)$. Consequently, after shifting the coordinates
by $x_1$, the NC Wightman functions will depend with different
functional forms on the two-dimensional vectors,
$\vec\sigma_i=(\xi^0_i,\xi^1_i)$ and $\vec\tau_i=(\xi^2_i,\xi^3_i)$,
\be\label{translation} W_\star(x_1,x_2,...,x_n)={\cal
W}_\star(\vec\sigma_1,\vec\tau_1,\vec\sigma_2,\vec\tau_2,...,\vec\sigma_{n-1},\vec\tau_{n-1})\
. \ee
The subscript $\star$ in the r.h.s. of \eqref{translation} shows
that the $\theta$-dependence of the Wightman functions is not
effaced by their translational invariance.

The translational invariance of NC Wightman functions  as in
\eqref{wightstar} was used as argument in \cite{FW} for the
disappearance of the $\star$-product upon translating by $\xi$,
since the $\star$-product of the four-dimensional shifts $\xi_i$
with any function of $x$ reduces to the usual product. We should,
however, be aware, that a function
\be\label{transl_4}F(x-y), \ \ \ \ x,y \in \mathbf R^{1,3}\ee
is translationally invariant, but so is also the function
\be\label{transl_2+2}F_0(x^0-y^0,\theta')F_1(x^1-y^1,\theta')F_2(x^2-y^2,\theta)F_3(x^3-y^3,\theta),\ee
with the notation used in \eqref{theta}. Indeed, the
$\star$-multiplication of two functions of the type \eqref{transl_4}
is the same like their usual multiplication, but the $\star$-product
of two functions of the type \eqref{transl_2+2} will retain its
$\theta$-dependence. The translational symmetry in the
relativistically invariant case falls into the first situation,
while in the noncommutative twisted Poincar\'e case, it falls into
the latter.

\subsection{Space of test functions for noncommutative Wightman
functions}\label{space_test_func}

Rigorously, field operators can not be defined at a point
\cite{Wightman} (see also Refs. \cite{WS} and \cite{BLT}), but only as smoothed
operators, written symbolically as
\begin{equation} \label{vff}
\vff \equiv\int \,\vfx \,\fsx \, d \, x,
\end{equation}
where $\fsx$ are test functions.

In QFT, the standard assumption is that all $\fsx$ are test
functions of tempered distributions. On the contrary, in the NC QFT
the corresponding generalized functions can not be tempered
distributions as the $\star$-product contains infinite number of
derivatives. It is well known (see, for example, Ref. \cite{WS}) that
there can be only a finite number of derivatives in any tempered
distribution.

In  Ref. \cite{test_funct} it was shown that the series
\begin{equation} \label{fsprodxy}
\fsx \star \fsy = \exp{\left ({\frac{i}{2} \, \theta^{\mu\nu} \,
\frac{ {\partial}}{\partial x^{\mu}} \, \frac{ {\partial}}{\partial
y^{\nu}}} \right)} \,\fsx  \fsy
\end{equation}
converges if $\fsx \in S^\beta, \; \beta <  1 / 2$, where $S^\beta$
is a Gel'fand-Shilov space \cite{GSh}. A similar result was obtained
also in Ref. \cite{MAS}. This space contains test functions
whose support is noncompact in the noncommutative directions, but is
compact in the commutative directions.

Thus the formal expression (\ref{wightstar}) actually means that the
scalar product of the vectors $\Phi_{k} = \vffkp |0\rangle$ and
$\Psi_n = \vffkpop |0\rangle$ is the following:
\begin{eqnarray}\label{scprod}
&&\langle \, \Phi_{k}, \Psi_n \, \, \rangle = \int\, \wf \, \fsbxsp
\dxo ... \dxn,\cr &&\wf
\,=\langle0|\varphi_\theta(x_1)...\varphi_\theta(x_n)|0\rangle.\end{eqnarray}
Let us stress that after integration over the noncommutative
variables $\wf$ becomes tempered distribution with respect to the
commutative variables $x_{i}^{0}, x_{i}^{1}, \, i = 1,2,..., n$.

We can use the formal expression (\ref{wightstar}) instead of the
rigorous one (\ref{scprod}). The point is that in accordance with
the spectral property, the Wightman functions are analytical
functions with respect to the commutative coordinates. This property
is crucial in  our proof of the CPT and spin-statistics theorems.
Let us point out that the CPT theorem and the spin statistics one can be
proved under conditions different from ours, namely by using the
concept of asymptotic commutativity condition \cite{Sol06}. We
assume that Wightman functions are tempered distributions with
respect to the commutative coordinates as this natural physical
assumption gives us the possibility to prove in NC QFT the main
axiomatic results of ordinary quantum field theory such as
irreducibility of the set of quantum field operators and generalized
Haag's theorem \cite{CMTV}.

\subsection{ Microcausality and spectral condition}\label{microcausality}

In order to be able to define a microcausality condition, we should
keep the locality in time, and therefore we shall restrict ourselves
to theories with space-space noncommutativity. We consider
henceforth $\theta'=0$ in \eqref{theta}. This situation corresponds
to a $O(1,1)\times{\cal T}_{(1,1)}={\cal P}(1,1)$ symmetry for the
$(x_0,x_1)$ plane and a $SO(2)\times{\cal T}_2=E_2$ symmetry for the
$(x_2,x_3)$ plane. According to the present wisdom, the postulate of
local commutativity (LC) is modified to require the vanishing of
star-commutators (imposed by the twisted-Poincar\'e
symmetry) of scalar fields at space-like separation in the sense of
$O(1,1)$ (i.e. replace light-cone by light-wedge):
\bn\label{lcc}[\phi_\theta(x),\phi_\theta(y)]_\star\equiv
\phi_\theta (x)\star \phi_\theta
(y)-\phi_\theta(y)\star\phi_\theta(x)=0,\cr
 \mbox{for}\ (x^0-y^0)^2-(x^1-y^1)^2<0.\en

The choice of the light-wedge microcausality condition
is based on the fact that the Moyal $\star$-product engenders
infinite nonlocality in the noncommutative coordinates, such that
the propagation of signal in the noncommutative directions is
instantaneous. Intuitively, it means that it "takes no time" for the
signal to propagate in those directions, therefore they drop out
from the separation of the causality-acausality regions. The result
is, as shown in Refs. \cite{SSB,LAG,test_funct,CNST} and references
therein, the enlargement of the light-cone to the light-wedge. We
should also point out that infinite nonlocality in time implied by a
noncommutative time coordinate defies quantization in the first
place, therefore this case can not be included in the study.

The results of Ref. \cite{test_funct} regarding the space of test
functions on which the $\star$-multiplication is well defined
reinforce the light-wedge microcausality condition -- these
functions have compact support in the commutative directions and
noncompact support in the noncommutative ones, the light-wedge
emerging thus naturally as domain of causality.

Concerning the local commutativity in the axiomatic formulation of NC QFT
with such and similar deformations as used in this paper, we should
mention that within the usual quantization prescription and the use of
perturbation theory, up to now there exists no specific model which
satisfies the LC property among the fields at the Jost points.

The lower symmetry compels us to enlarge, correspondingly, the
physical spectrum of the momentum operator, i.e. the spectral
condition, to the forward light-wedge:
\be\label{spectrum}Spec\, (p)=\{(p^0)^2-(p^1)^2\geq 0,\ p^0\geq0\}\
. \ee

On the ground of this formulation we shall prove the CPT theorem for
theories with space-space \ncy and also the spin-statistics theorem
for the simplest case of scalar field.

\section{CPT theorem in the axiomatic approach to NC QFT}\label{cpt-theorem}

In ordinary QFT, CPT theorem was proven in the Lagrangian approach,
by L\"uders and Pauli \cite{Luders}. A general proof was given by
Jost in the axiomatic formulation \cite{jost-cpt} (see also Ref.
\cite{WS}). In the NC case, CPT theorem was shown to hold in NC QED
\cite{Shahin}. A general proof in the Lagrangian formalism, for any
NC QFT, was given in Ref. \cite{cnt}.

In the axiomatic approach to the ordinary QFT, the CPT theorem
states \cite{WS} that the CPT invariance condition in terms of
Wightman functions, e.g. in the case of a neutral scalar field,
\be\label{cpt-inv}W(x_1, x_2,...,x_n)=W(-x_n,...,-x_2,-x_1)\ , \ee
for any values of $x_1$, $x_2$,...,$x_n$, is equivalent to the weak
local commutativity (WLC) condition, \be\label{micro}W(x_1,
x_2,...,x_n)=W(x_n,...,x_2,x_1)\ , \ee where  $x_1-x_2$,
...,\,$x_{n-1}-x_{n}$ is a Jost point, i.e. it satisfies the
condition that
$$\left(\sum\limits_{j=1}^{n-1}\lambda_j(x_{j}-x_{j+1})\right)^2<0,\
\ \mbox{for all}\ \  \lambda_j\geq0 \ \ \mbox{with}\ \
\sum\limits_{j=1}^{n-1}\lambda_j>0.$$
The main ingredients for the proof are the analyticity of the
Wightman functions and the fact that the space-time inversion
(PT-transformation) is connected to the identity in the complex
Lorentz group.

\subsection{ CPT invariance and WLC conditions in terms of the NC
Wightman functions}

In order to prove the CPT theorem in NC QFT, the first of our
concerns is to derive the CPT invariance in terms of the new
Wightman functions defined by eq. (\ref{wightstar}).

In the following, for simplicity, we shall restrict ourselves to the
case of one neutral scalar field.

The CPT invariance condition is derived by requiring that the CPT
operator $\Theta$ be antiunitary (see, e.g. \cite{WS}):
\be\label{antiunit}\langle\Theta\Phi|\Theta\Psi\rangle=\langle\Psi|\Phi\rangle\
, \ee i.e. the CPT operator leaves invariant all transition
probabilities of the theory.

Taking the vector states as
$\langle\Phi|=\langle0|\equiv\langle\Psi_0|$ and
$|\Psi\rangle=\phi_\theta(x_n)\star...\star\phi_\theta(x_2)\star\phi_\theta(x_1)|\Psi_0\rangle$
we shall express both sides of (\ref{antiunit}) in terms of NC
Wightman functions.

For the l.h.s. of (\ref{antiunit}) we use directly the CPT
transformation properties of the field operators, which read, for a
neutral scalar field,
$\Theta\phi_\theta(x)\Theta^{-1}=\phi_\theta(-x)$. Using the
CPT-invariance of the vacuum state,
$\Theta|\Psi_0\rangle=|\Psi_0\rangle\equiv|0\rangle$, the l.h.s. of
(\ref{antiunit}) becomes:
\bn\label{thetaw2}
\langle\Theta\Phi|\Theta\Psi\rangle&=&\langle\Theta\Psi_0|\Theta(\phi_\theta(x_n)\star...\star\phi_\theta(x_2)\star\phi_\theta(x_1)|\Psi_0\rangle)\cr
&=&W_{\star}(-x_n,...,-x_2,-x_1)\ . \en
For expressing the r.h.s. of (\ref{antiunit}) we take the hermitian
conjugates of the vectors $|\Psi\rangle$ and $\langle\Phi|$, to
obtain: \be\label{thetaw1}
\langle\Psi|\Phi\rangle=W_\star(x_1,x_2,...,x_n)\ . \ee
Putting together (\ref{antiunit}) with (\ref{thetaw2}) and
(\ref{thetaw1}), we obtain the CPT invariance condition in terms of
NC Wightman functions as \be\label{cpt-wight}
W_\star(x_1,x_2,...,x_n)=W_{\star}(-x_n,...,-x_2,-x_1)\ .\ee

 Let us introduce the WLC condition. We remark
that the $\star$-products contained in the definition of the
Wightman functions do not influence in any way the coordinates
involved in defining the light-wedge in (\ref{lcc}), i.e. $x^0$ and
$x^1$. Consequently, at space-like separated points in the sense of
$SO(1,1)$ (denoted by $x_i\sim x_j$, $i,j=1,2,...,n$), we can
permute the field operators in (\ref{wightstar}) in accordance with
(\ref{lcc}). The WLC condition implies only that Wightman functions
are not changed if the direct order of points $x_{i}$ is substituted
by the inverse one. Thus the \nc version of the WLC condition in
terms of Wightman functions reads: \bn\label{wlc} W_\star(x_1,
x_2,...,x_n)&=&W_{\star}(x_n,...,x_2,x_1)\,\cr \mbox{for} \ \
x_i&\sim& x_{j}\ , \ i,j=1,2,...,n\ . \en

Finally the proof of the CPT theorem amounts to showing the
equivalence of (\ref{cpt-wight}) and (\ref{wlc}).

\subsection{Proof of CPT theorem}

The CPT theorem consists of the equivalence of (\ref{cpt-wight}) and
(\ref{wlc}). The proof goes along similar lines as in the
commutative case \cite{WS} or the case discussed in Ref. \cite{LAG}. The
main step is the analytical continuation of the Wightman functions
to the complex plane only with respect to $x_0$ and $x_1$. The
$\star$-products introduced in the new Wightman functions have no
influence on these two coordinates (since we have taken
$\theta_{01}=0$, i.e. the time to be commutative, cf. \eqref{theta})
and upon analytical continuation they will not be affected. Due to
the translational invariance, we can express $W_\star(x_1,
x_2,...,x_n)$ in terms of the $2(n-1)$ relative variables,
$\vec\sigma_i=(\xi^0_i,\xi^1_i)$ and $\vec\tau_i=(\xi^2_i,\xi^3_i)$,
with $\xi^\mu_i=x^\mu_i-x^\mu_{i+1}$:
\be\label{transl} W_\star(x_1,x_2,...,x_n)={\cal
W}_\star(\vec\sigma_1,...,\vec\sigma_{n-1},\vec\tau_1,...,\vec\tau_{n-1})
\ee
(see the discussion in Section 2.4).

From (\ref{spectrum}), the spectral condition for NC Wightman
functions follows. Specifically, the Fourier transform of a Wightman
function is nonzero, i.e.
 \bn\label{spectr-cond} \tilde {\cal
W}_\star(\vec{\cal P}_1, \vec{\cal P}_2,...,\vec{\cal
P}_{n-1})=\frac{1}{(2\pi)^{2(n-1)}}
\int\Pi_{i=1}^{n-1}d\vec\sigma_k\cr \times\ e^{-i({\cal
P}^0_k\sigma_k^0-{\cal P}^1_k\sigma_k^1)} {\cal
W_\star}(\vec\sigma_1,...,\vec\sigma_{n-1},\vec\tau_1,...,\vec\tau_{n-1})\neq
0\ . \en
only if all the two-dimensional momenta $\vec{\cal P}_i=({\cal
P}_i^0,{\cal P}_i^1)$ satisfy the conditions (\ref{spectrum}). In
(\ref{spectr-cond}) the inessential dependence of the l.h.s. on the
vectors $\vec\tau_i$ is omitted. The condition (\ref{spectr-cond})
implies that, on the same grounds as in commutative case \cite{WS},
NC Wightman functions can be analytically continued into the complex
plane with respect to $\xi^0_i$ and $\xi^1_i$, by the substitution
$\xi_i\rightarrow\mu_i=\xi_i-i\eta_i$, with all $\eta_i$ belonging
to the forward light-wedge, i.e. $(\eta_i^0)^2-(\eta_i^1)^2\geq0, \
\ \eta_i^0\geq0$ and $\eta_i^2=\eta_i^3=0$. In other words,
\bn\label{fourier} {\cal W}_\star(\mu_1,
\mu_2,...,\mu_{n-1})=\frac{1}{(2\pi)^{2(n-1)}}\int\Pi_{k=1}^{n-1}
d\vec{\cal P}_k\cr \times\ e^{i({\cal P}^0_k\mu_k^0-{\cal
P}^1_k\mu_k^1)} \tilde {\cal W}_\star(\vec{\cal P}_1, \vec{\cal
P}_2,...,\vec{\cal P}_{n-1}) \en is analytical in $\mu_i^0$ and
$\mu_i^1$. In accordance with the Bargmann-Hall-Wightman theorem
\cite{WS,BHW}, the functions ${\cal W}_\star(\mu_1,
\mu_2,...,\mu_{n-1})$ are analytical in an extended domain; in
commutative case the extended domain is obtained from the initial
one by applying all (proper) complex Lorentz transformations,
continuously related to the unit transformation. In the \nc case we
can use only the transformations belonging to the complex $O(1,1)$
\cite{LAG} for obtaining the extended domain of analyticity. This
domain contains also the real points $\tilde\xi_i$ called Jost
points, with the property that
$\left(\sum\limits_{j=1}^{n-1}\lambda_j\tilde\xi^0_{j}\right)^2-
\left(\sum\limits_{j=1}^{n-1}\lambda_j\tilde\xi_{j}^1\right)^2<0$,
for all $\lambda_j\geq0$ with $\sum\lambda_j>0$ (consequently, the
points $x_1$, $x_2$,...,$x_n$ are mutually space-like in the sense
of $O(1,1)$). The values of Wightman functions at Jost points fully
determine the values in the whole domain of analyticity. Thus, two
Wightman functions coinciding at their Jost points coincide
everywhere.

We shall show now that the WLC condition (\ref{wlc}) implies the CPT
invariance condition (\ref{cpt-wight}). Let us first rewrite
(\ref{wlc}) in terms of relative coordinates: \be\label{wlc'} {\cal
W}_\star(\tilde\xi_1, \tilde\xi_2,...,\tilde\xi_{n-1})={\cal
W}_{\star}(-\tilde\xi_{n-1},...,-\tilde\xi_2, -\tilde\xi_1)\ . \ee
The functions ${\cal W}_\star(\xi_1,...,\xi_{n-1})$ and ${\cal
W}_{\star}(-\xi_{n-1},...,-\xi_1)$ satisfy the spectral condition
(\ref{spectr-cond}) and are invariant under $O(1,1)$
transformations. Thus, in accordance with the previous arguments,
they are both analytical functions of the complex variables $\mu_i$
in the above-mentioned extended domain. Moreover, since they are
equal at the Jost points, they are also equal in the whole domain of
analyticity. Using the invariance of ${\cal W}_\star(\mu_1,
\mu_2,...,\mu_{n-1})$ and ${\cal
W}_{\star}(-\mu_{n-1},...,-\mu_2,-\mu_1)$ under the complex $O(1,1)$
group, which includes the inversion $\mu_i^0\rightarrow-\mu_i^0$ and
$\mu_i^1\rightarrow-\mu_i^1$, we arrive at the equality
\be {\cal W}_\star(\mu_1,...,\mu_{n-1})={\cal
W}_{\star}(-\mu'_{n-1},...,-\mu'_1)\ , \ee
where
$\mu'_i=(-\mu_i^0,-\mu_i^1,\mu_i^2,\mu_i^3)\equiv(-\mu_i^0,-\mu_i^1,\tau_i^2,\tau_i^3)$.
Performing a $SO(2)$ rotation by $\pi$ in the ($\tau^2,\tau^3$)
plane and subsequently going to the real limit, we obtain that \be
{\cal W}_\star(\xi_1,\xi_2,...,\xi_{n-1})={\cal
W}_{\star}(\xi_{n-1},...,\xi_2, \xi_1)\ , \ee which is equivalent to
the CPT invariance condition (\ref{cpt-wight}) in terms of $x_1$,
$x_2$,...,$x_n$. Thus CPT invariance is the consequence of WLC. By
similar considerations the converse can also be proven, i.e. CPT
invariance implies WLC.

\section{Spin-statistics theorem}

Within the Lagrangian framework the spin-statistics theorem has been
shown \cite{cnt} to hold for NC QFT with space-space
noncommutativity. The proof used Pauli's original formulation
\cite{Pauli}, requiring that at space-like separations the
commutators of two observables should vanish.

Moreover, the symmetry under twisted Poincar\'e algebra gives an
additional hint that the spin-statistics relation should survive in
the NC QFT. In the case of a usual Lie algebra, the operator of
permutation $P$ inverts the order of two representations in a tensor
product: $P(a\otimes b)=b\otimes a$. In the case of a quantum group
with a universal enveloping algebra $\cal R$, the notion of
permutation operator changes, such that the new permutation
operator, $\Psi$, is consistent with the quantum group action, i.e.
$\Delta (Y)(\Psi(a\otimes b))=\Psi(\Delta (Y)(a\otimes b))$, where
$\Delta(Y)$ is the deformed coproduct of a Lie algebra generator.
Using \eqref{r-matrix}, it follows that $\Psi(a\otimes b)=P({\cal
R}(a\otimes b))$. The twisted Poincar\'e algebra is a particular
case of quantum group, called triangular Hopf algebra, for which
${\cal R}^{-1}={\cal R}_{21}$. Consequently, $\Psi=\Psi^{-1}$, i.e.
$\Psi$ is symmetric and no exotic statistics emerges, but the
representations of the twisted Poincar\'e algebra will have the same
statistics as those of the ordinary Poincar\'e symmetry (see Refs. 
\cite{monographs}). These aspects refer to the spin-statistics
\emph{relation} for particle representations, and not to quantum
field theory. Similar considerations based on the twisted Poincar\'e
symmetry can be found as well in Ref. \cite{spinstat}.

Encouraged by these indications, we attempt here to show the
validity of the spin-statistics theorem based on the NC Wightman
functions, for the case of a real scalar field.

We start by proving that, if $\psi_\theta(x)|0\rangle=0$ and
$\psi_\theta(x)$ is a local field operator in the sense of the
light-wedge, then $\psi_\theta(x)=0$. To show this, we take at the
Jost points $\tilde x_1-\tilde x_2$,...,$\tilde x_j-\tilde x$,
$\tilde x-\tilde x_{j+1}$ ,...,$\tilde x_{n-1}-\tilde x_n$ the
arbitrary NC Wightman function \bn\label{lemma}
&&\langle0|\phi_\theta(\tilde x_1)\star...\star\phi_\theta(\tilde
x_{j})\star\psi_\theta(\tilde x)\star\phi_\theta(\tilde
x_{j+1})...\star\phi_\theta(\tilde x_n)|0\rangle\cr
&&=\langle0|\phi_\theta(\tilde x_1)...\phi_\theta(\tilde
x_{j})\star\phi_\theta(\tilde x_{j+1})...\phi_\theta(\tilde
x_n)\star\psi_\theta(\tilde x)|0\rangle=0\ . \en
By analytically continuing the first line of (\ref{lemma}) (in
exactly the same way as for the proof of the CPT theorem in the
Section \ref{cpt-theorem}), we obtain
$\langle0|\phi_\theta(x_1)\star...\star\phi_\theta(x_{j})\star\psi_\theta(x)
\star\phi_\theta(x_{j+1})...\star\phi_\theta(x_n)|0\rangle=0$, i.e.
all the matrix elements of the operator $\psi_\theta(x)$ between a
complete set of states
$\langle0|\phi_\theta(x_1)\star...\star\phi_\theta(x_{j})$ and
$\phi_\theta(x_{j+1})...\star\phi_\theta(x_n)|0\rangle$ are zero and
thus $\psi_\theta(x)=0$.

Now we show that the wrong statistics, \be\label{wrong}
\{\phi_\theta(x),\phi_\theta(y)\}_\star=0,\
(x^0-y^0)^2-(x^1-y^1)^2<0\ , \ee leads to $\phi_\theta(x)=0$.

Consider
$W_\star(x,y)=\langle0|\phi_\theta(x)\star\phi_\theta(y)|0\rangle$.
According to (\ref{wrong}) we have \be\label{5} W_\star(\tilde
x,\tilde y)+ W_{\star}(\tilde y,\tilde x)=0\ . \ee Eq. (\ref{5}) can
be analytically continued, as in the Section \ref{cpt-theorem}, into the
extended domain. Performing a space-time inversion and taking the
real limit for the coordinates, we obtain for the second term of
(\ref{5}) $ W_{\star}(y,x)= W_{\star}(x,y)$. Thus, $W_\star(x,y)=0$.
At $y=x$ we get \be\label{norm}
\langle0|\phi_\theta(x)\star\phi_\theta(x)|0\rangle=0\ , \ee which
is equivalent to $\langle\Psi|\Psi\rangle=0$, with
$|\Psi\rangle=\phi_\theta(x)|0\rangle$, if one adopts the definition
for the norm of a state as in (\ref{norm}), or equivalently as
$\langle\Psi|\Psi\rangle=\langle0|\phi_\theta(\hat
x)\phi_\theta(\hat x)|0\rangle$. Then $\phi_\theta(x)|0\rangle=0$
and, due to the result first derived, we get $\phi_\theta(x)=0$.

\section{Discussion and conclusions}

In this paper we have introduced new Wightman functions defined as
vacuum expectation values of the product of field operators with
$\star$-multiplication. In this attempt to develop an axiomatic formulation
of NC QFT, the \ncy of space-time is explicitly taken into account.
Using such NC Wightman functions, for the case of only space-space
noncommutativity, \ss, we prove the CPT theorem adapted to the
symmetry \group. The results presented in  this paper show that CPT invariance can be valid for theories
which do not possess Lorentz symmetry. Conversely, Lorentz-invariant but
CPT-violating (nonlocal) quantum field theories can also be constructed
\cite{CDNT}. Thus, CPT invariance and Lorentz symmetry can be disentangled, at
least for nonlocal quantum field theories. 

In a quite different context, the description of Nature
at the Planck scale is suggested to be given by a nonlocal
translationally invariant theory, the so-called Very Special
Relativity (VSR) \cite{VSR}, with a symmetry under a subgroup of the
Lorentz group, while at low-energy scale the Poincar\'e invariance
would be operating. The CPT symmetry is considered to work
throughout the energy scales, while the violation of P, T and CP
symmetries is connected to the violation of Lorentz symmetry. A
robust mathematical framework for VSR has been recently proposed
within noncommutative quantum field theory with light-like
noncommutativity \cite{NCVSR}. The present results on CPT invariance
and spin-statistics relation carry over to light-like
noncommutativity as a realization of VSR.

 For the case of a scalar field we have also
proven the spin-statistics theorem using the NC Wightman functions
with \ss.

The twisted Poincar\'e symmetry of NC QFT is crucial for the
accurate axiomatic formulation: {\it i)} it justifies the use of the
$\star$-product in the construction of the Wightman functions (see
Section \ref{NCwight}) and {\it ii)} it allows us altogether to
speak rigorously about the concept of spin in NC QFT (a concept
which would be inexistent in the residual symmetry formulation) and
thus to pose the problem of the spin-statistics relation (see
Section \ref{twistPs} and Ref. \cite{CKNT}). For the actual proof of
the CPT and spin-statistics theorem, the interplay of the twisted
Poincar\'e and the residual $O(1, 1)\times SO(2)$ symmetry is
essential (see Section \ref{NC_field_op} and Refs.
\cite{CKTZZ,CNST}). The analytical properties of the smeared new
Wightman functions can be utilized in the proofs due to the
existence of the corresponding class of test functions, namely the
Gel'fand-Shilov space $S^\beta$, $\beta < 1/2$ (see Section
\ref{space_test_func} and Ref. \cite{test_funct} for details).
The causality condition used is the vanishing of
star-commutator of the scalar field at light-wedge space-like
points, as in eq. (\ref{lcc}), which is fully justified only in the
context of the Gel'fand-Shilov space $S^\beta$, $\beta < 1/2$, whose
functions have compact support only in the commuting directions.

There are still several important questions to be studied within the
axiomatic formulation of  NC QFT, e.g. the cluster decomposition
property, the uniqueness of the vacuum state related to the
irreducibility of field operator algebra, the proof of the analytic
properties of the scattering amplitude \cite{Sibold} and their
implication on its high-energy behaviour \cite{JLD}. Finally, among
the most important problems remains the formulation and proof of the
reconstruction theorem. We hope to present further results on these
problems in a future communication.

\section*{Acknowledgements} We are much grateful to L.
\'Alvarez-Gaum\'e and C. Montonen for useful discussions. The
support of the Academy of Finland under the Projects No. 136539 and 140886 is gratefully acknowledged.

\end{document}